\documentclass[aps,preprint,amssymb,12pt,floatfix]{revtex4}
\usepackage{subfigure}
\usepackage{graphicx}
\usepackage{rotating} \usepackage{color}
 \usepackage{amsmath,amsthm}
\usepackage{epstopdf}
\begin{document}

\title{Using Simulations and kinetic network models to reveal the dynamics and functions of Riboswitches}


\author{Jong-Chin Lin$^{1}$, Jeseong Yoon$^{2}$, Changbong Hyeon$^{2}$, and D. Thirumalai$^{1}$}
\affiliation{School of Computational Sciences, Korea Institute for Advanced Study, Seoul 130-722}
\affiliation{Institute for Physical Science and Technology and Department of Chemistry and Biochemistry, University of Maryland, College Park 20742, USA}

\date{\today}

\begin{abstract}
Riboswitches, RNA elements found   in the untranslated region, regulate gene expression by binding to target metaboloites with exquisite specificity.  Binding of metabolites to the conserved aptamer domain allosterically alters  the conformation in the downstream expression platform.  The fate of gene expression is determined by the changes in the downstream RNA sequence. 
As the metabolite-dependent cotranscriptional folding and unfolding dynamics of riboswitches is the key determinant of gene expression, it is important to investigate both the thermodynamics and kinetics of riboswitches both in the presence and absence of metabolite. 
Single molecule force experiments that decipher the free energy landscape of riboswitches from their mechanical responses, theoretical and computational studies have recently shed light on the distinct mechanism of folding dynamics in different classes of riboswitches.
Here we first discuss the dynamics of water around riboswitch, highlighting that water dynamics can enhance the fluctuation of nucleic acid structure.   To go beyond native state fluctuations we used the
Self-Organized Polymer (SOP) model to predict the dynamics of add adenine riboswitch under mechanical forces. In addition to quantitatively predicting the folding landscape of add-riboswitch our simulations also explain the difference in the dynamics between pbuE adenine- and add adenine-riboswitches.  In order to probe the function {\it in vivo} we use the folding landscape to propose a system level kinetic network model to quantitatively predict how gene expression is regulated for riboswitches that are under kinetic control.
\end{abstract}

\maketitle
\section{Introduction and Scope of the Review}

The remarkable discovery just over a decade ago that riboswitches, which are RNA elements in the untranslated regions in mRNA, control gene expression by sensing and binding target metabolites with exquisite sensitivity is another example of the versatility of RNA in controlling crucial cellular functions \cite{Serganov13Cell}. In the intervening time, considerable insights into their functions have come from a variety of pioneering biochemical and biophysical experiments. In addition, determination of structures of a number of riboswitches has greatly aided in molecular understanding of their functions and in the design of synthetic riboswitches. Typically, riboswitches contain a conserved aptamer domain to which a metabolite binds, producing a substantial conformational change in the downstream expression platform leading to control of gene expression. The variability in the functions of structurally similar aptamer domain is remarkable. The {\it add} adenine (A) 
riboswitch activates translation upon binding the metabolite (purine) whereas the
structurally similar {\it pbuE} adenine (A) riboswitch controls
transcription \cite{Mandal03Cell}. We can classify both these as ON riboswitches,
which means that translation
or transcription is activated only upon binding of the metabolite.
In contrast, OFF riboswitches (for example Flavin Mononucleotide (FMN) binding aptamer)  shut down gene expression
when the metabolite binds to the aptamer domain. 
Finally, some of the riboswitches are under kinetic control (FMN riboswitch \cite{Wickiser2005MolCell} and
{\it pbuE} adenine riboswitch \cite{Frieda2012Science}) whereas others
(for example {\it add} adenine riboswitch and SAM-III riboswitch) may be under thermodynamic control. 

The functions of riboswitches are vastly more complicated than indicated by {\it in vitro} studies, which typically focus on limited aspects of their activities. Under cellular conditions the metabolite which binds to the aptamer domain itself is a product of gene expression. Thus, regulation  of gene expression involves negative or positive (or a combination) feedback. This implies that a complete understanding of riboswitch function must involve a system level description, which should minimally include the machinery of gene expression, rate of transcription or translation, degradation rates  of mRNA, and activation rate of the synthesized metabolite as well as rate of binding (through feedback loop) to the aptamer domain. Many {\it in vitro} studies have dissected these multisteps into various components in order to quantify them as fully as possible. In this context, single molecule studies in which response of riboswitches to mechanical force are probed have been particularly insightful \cite{Greenleaf08Science,Frieda2012Science}. 

More recently, computational and theoretical studies have been initiated to develop a quantitative description of the folding landscape and dynamics at the single molecule level \cite{Lin08JACS,whitford2009nonlocal,Allner13RNA,Quarta12PlosCompBiol,Feng11JACS}. The findings have been combined to produce a framework for describing the function of riboswitches at the system level. Because atomically detailed simulations can only provide limited information of the dynamics in the folded aptamer states it is necessary to develop suitable coarse grained (CG) model for more detailed exploration of the dynamics. The CG models of nucleic acids, first introduced by Hyeon and Thirumalai \cite{Hyeon05PNAS}, are particularly efficacious to deal with riboswitches that undergo large scale conformational fluctuations for functional purposes. The goal of this article is limited to a brief review of the insights molecular simulations have brought to the understanding of the folding landscape of riboswitches using purine-binding and S-adenosylmethionine (SAM) riboswitches as examples. We begin with the description of the fluctuations in the folded state of the small preQ1 riboswitch, which exhibits rich dynamics in the native state that can be probed using atomic detailed simulations in explicit water. The hydration dynamics could have a functional role, which can be resolved by spectroscopic experiments.   We then describe the response of three classes of riboswitches to forces and map the entire folding landscape from which we have made testable predictions.  The data from these free energy landscapes are used to construct a network model, which provides system level description of the OFF riboswitches (Fig.\ref{ribo}). 

\section{Hydration dynamics around the folded state: All atom simulations}
It is known that unlike proteins there are many are several low free energy excitations (alternate structures) that a folded RNA can access. Consequently, dynamical fluctuations of the folded  states are critical for RNA to execute their biological functions. 
There is growing evidence that hydration plays a key role in triggering conformational fluctuations in RNA. First, RNA can access low-lying excitation states via local melting of bases 
\cite{jacob1987PNAS,Dethoff2012Nature}. 
Recent NMR studies suggest that a potential pre-melting of the hydration shell is required for the base pair disruption in response to elevation in temperature \cite{Rinnenthal10NAR,Nikolova10RNA}. 
Second, the versatile functional capacity of RNAs can be attributed to their ability to access alternative conformations 
\cite{Zhang06Science}. 
Local conformational fluctuations from a few nanosecond dynamics enable RNA to explore a heterogeneous conformational ensemble, giving  them the capacity to recognize and bind a diverse set of ligands. 
Third, binding of metabolites to riboswitches to control gene expression \cite{Montange08ARB} may also be linked to local fluctuations in specific regions of co-transcriptional folded UTR regions of mRNA. In all of these examples, hydration of RNA is likely to play important role.

In order to illustrate the importance of hydration we performed atomically detailed simulations of PreQ$_1$ riboswitch. We showed that water dynamics is spatially heterogeneous with metastable functionally relevant states whose dynamics and spans many orders of magnitude. This behavior is reminiscent of glassy behavior \cite{Thirumalai89PRA}. 
The glassy behavior of water molecules may indicate that RNA molecule is able to access low-lying free energy states around the putative folded state. We identified  distinct classes of water molecules near the RNA surface, which can be classified as ``bulk", ``surface", ``cleft", and ``buried" water in the order of increasing water hydrogen bond relaxation time \cite{yoon2014JPCB}. 
In this section we review the molecular details of hydration around various regions of RNA and discuss how water dynamics gives rise to local structural fluctuations by using atomic simulations of PreQ$_1$-riboswitches \cite{Yoon13JACS,yoon2014JPCB}.   
\\

{\it  Water hydrogen bond kinetics around nucleotides.} 
The time- and ensemble-averaged auto-correlation function $c(t)$ (see definition in Ref.\cite{yoon2014JPCB,Luzar96PRL}) are used to quantify the structure and dynamics of water molecules near the surface of RNA. 
The relaxation kinetics of the water HB at T=310 K around three different nucleotide groups (B: base, R: ribose, P: phosphate) is fitted to a multiexponential function $c_{\xi}(t)=\sum_{i=1}^N\phi_ie^{-t/\tau_i}$  where $\sum_{i=1}^N\phi_i=1$ with $N=4$ and $\xi$ denotes B, P, or R (Fig.\ref{hydration}a, left). 
The time constant $\tau_i$ ranges from $\mathcal{O}(1)$ ps to $\mathcal{O}(10^4)$ ps, but 90 \% of kinetics is described by the dynamics of $\lesssim\mathcal{O}(10^2)$ ps (Fig.\ref{hydration}a, left). 
The average lifetime of the water HB at each nucleotide group ($\langle\tau_{\xi}\rangle=\int^{\infty}_0dtc_{\xi}(t)$) reveals that water molecules exhibit the slowest relaxation dynamics near bases instead of phosphate groups as  $\langle\tau_P^w\rangle$= 193 ps, $\langle\tau_R^w\rangle$= 81 ps, $\langle\tau_B^w\rangle$ = 289 ps, and hence $\langle\tau_R^w\rangle<\langle\tau_P^w\rangle<\langle\tau_B^w\rangle$ .  It is worth pointing out that these time scales are much longer than found in proteins, and far exceed by a few orders of magnitude hydrogen bond dynamics in bulk water. 

Excess monovalent counter-ions are distributed around RNA to neutralize the negative charges on the phospho-diester backbone of nucleic acids. The auto-correlation functions computed for Na$^+$ ions bound to P, R, and B (Figure \ref{hydration}a, right) show that the time constant of ion relaxation is a few orders of magnitude greater than the water hydrogen bonds, $\langle\tau_P^{\text{Na}^+}\rangle$ = 294 ns, $\langle\tau_R^{\text{Na}^+}\rangle$ = 63 ns, $\langle\tau_B^{\text{Na}^+}\rangle$ = 9.2 ns. The order of lifetime differs from that of water HB as $\langle\tau_B^{\text{Na}^+}\rangle<\langle\tau_R^{\text{Na}^+}\rangle<\langle\tau_P^{\text{Na}^+}\rangle$. In contrast to water HB, monovalent counterions have the slowest dynamics near the phosphate group. 
Most importantly, while binding or release of a Na$^+$ ion to or from the surface of RNA certainly perturbs the water environment  \cite{Song2014JACS}, the time scale separation between water and counterion dynamics ensures that the hydration dynamics around RNA occurs essentially in a static ionic environment. 
\\

{\it Heterogeneity of Water Dynamics on the RNA Surface.}
The time scale of hydrated water varies many orders of magnitude depending on its location on the surface of RNA. 
Calculations of electrostatic potential on the solvent accessible surface confirm \cite{yoon2014JPCB,Yoon13JACS} that the charge distribution on RNA surface is indeed not uniform but heterogeneous. 
Multi-exponential function $c(t)=\sum_{i=1}^N\phi_ie^{-t/\tau_i}$ with different weights ($\phi_i$) and well-separated time constants ($\tau_i$) is needed to quantitate the relaxation dynamics of water molecules around four selected nucleotides of preQ$_1$-riboswitch, 24U, 29A, 33C, and 35A (Fig.\ref{hydration}c). The rich dynamics  
reflects the heterogeneity and justifies the interpretation that there are distinct class of water molecules, which can be  divided into multiple classes such as  ``bulk", ``surface", ``cleft", and ``buried" water \cite{yoon2014JPCB}.
At high temperatures, the population of fast, bulk water-like dynamics is dominant, but as the temperature decreases, the population of slow dynamics grows.   
The average lifetime of water molecules near RNA is at least 1--2 orders of magnitude slower than that of bulk water over the broad range of temperatures (Fig.\ref{hydration}c). 
\\

{\it Water-induced fluctuations of base-pair dynamics.} 
Dynamic feature of water that induce local conformational fluctuation of RNA is captured by probing the base pair dynamics along with surface water \cite{Yoon13JACS}. 
The space made of base stacks and base pairings is generally dry and hydrophobic, and thus devoid of any water molecules.  
However,in base pairs located at the end of stacks, it is possible to observe an enhanced fluctuation of base pair. 
Figure \ref{hydration}c shows the dynamics of base pairs A3-U24 located in the 5'- and 3'-end in preQ$_1$ riboswitch in aqueous solution. 
Remarkably, when the time series of water density around H3 of U24 and breathing dynamics of the base pair are compared, the change in water density always precedes the change in base-pair distance. 
The water densities calculated in the first and second solvation shell around H3 of AU24 show that water population starts to increase before the base pair disruption; the decrease of water population always precedes the event of base-pair formation. 
Thus, we conclude that the dynamics of water hydration and dehydration induces the breathing dynamics of base pairs. The spontaneous fluctuations in base pair opening induced by water are important in protein-DNA interactions as well and may be responsible for transcription initiation by RNA polymerase.

\section{Stability of isolated helices control the folding landscapes of purine riboswitches}
\begin{figure}
\includegraphics[width=1.0\columnwidth]{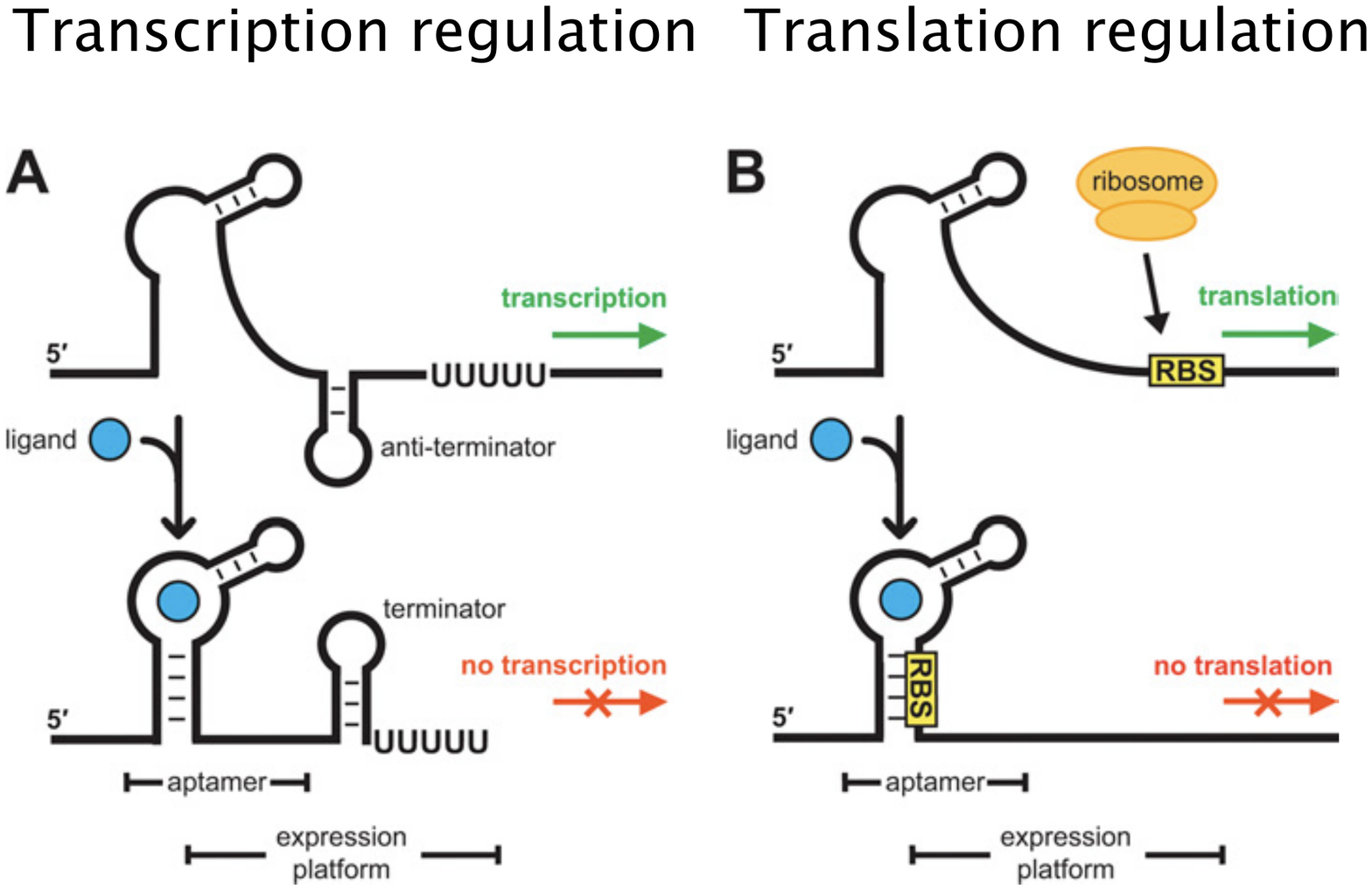}
\caption{Principles of transcription and translation regulation by OFF-riboswitches.  (A) Metabolite binding to the aptamer domain results in the formation of the terminator hairpin exposing a stretch of Uracil nucleotides. The polymerase disengages from the gene, thus terminating transcription. (B) Binding of the metabolite encrypts the Ribosome Binding Site (RBS) preventing the ribosome to initiate translation. The figure adapted from Ref.\cite{Kim2008BCell}.
\label{ribo}}
\end{figure}

\begin{figure}
\includegraphics[width=0.8\columnwidth]{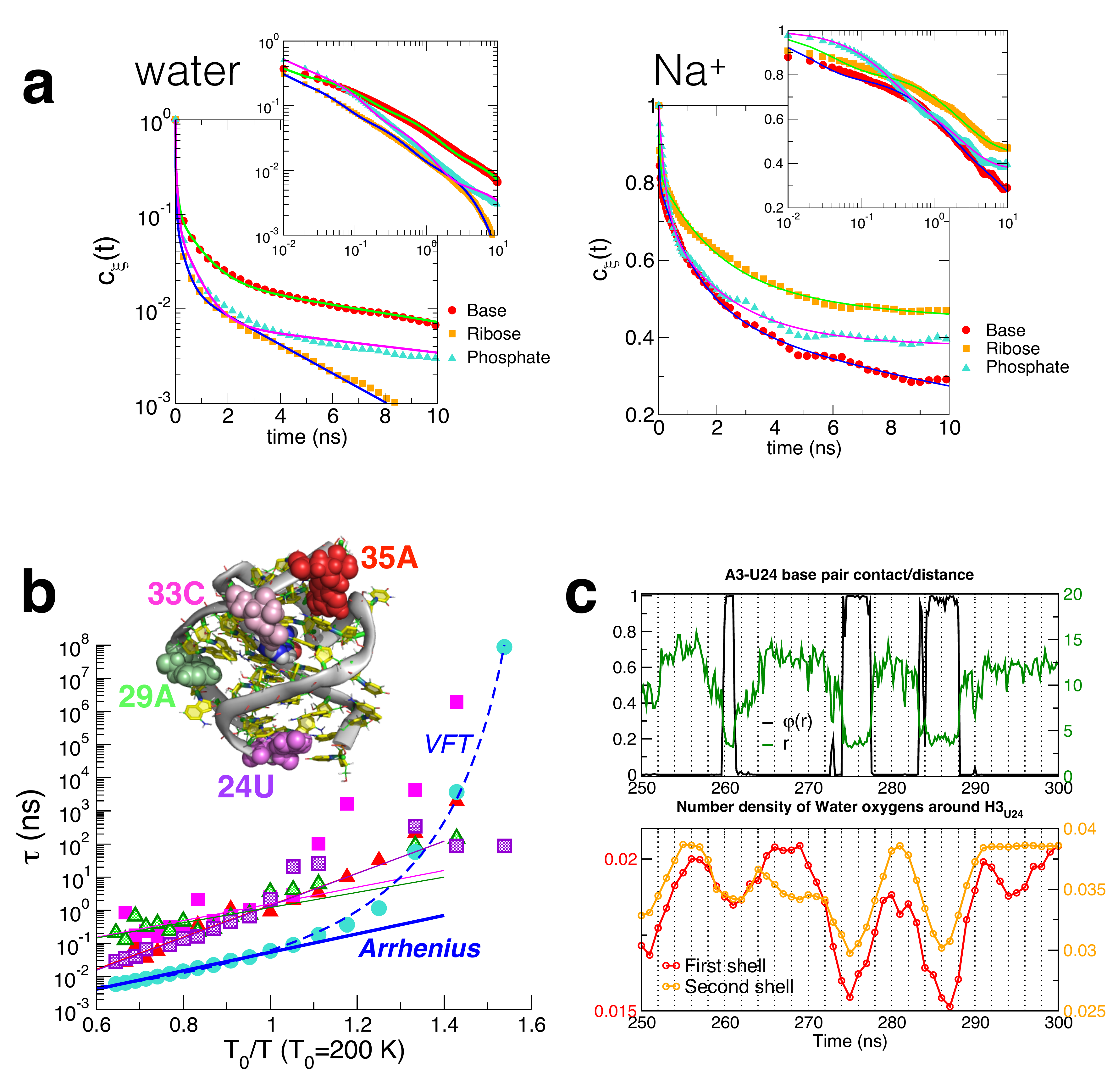}
\caption{a. The data for the dynamics of water HBs (left) and Na$^+$ (right) with base, ribose, and base (left) are described by the appropriate auto-correlation functions. b. Temperature dependence of water HB relaxation time of bulk water and around four nucleotides, 24U, 29A, 33C, and 35A or PreQ$_1$. 
Arrhenius fits are made to $T>200$ K for the four nucleotide and bulk water. 
Alternatively, the relaxation dynamics of bulk water HB can be fit using  the Vogel-Fulcher-Tamman (VFT) equation over the full range of temperatures (dashed line) \cite{yoon2014JPCB}. 
c. Water dynamics induced fluctuations of a base pair. 
The status of base pair (N1$_{A3}$--N3$_{U24}$), quantified by calculating logistic function $\varphi(r)=(1+e^{(r-r_0)/\sigma})^{-1}$ as well as distance $r$, shows apparent fluctuations, whose time scale is $\sim$ 10 ns.
Figures adapted from Refs.\cite{Yoon13JACS,yoon2014JPCB}
\label{hydration}}
\end{figure}

A key event in the function of riboswitches is the conformational change in the aptamer domain leading to the formation of the terminator with the downstream expression platform (Fig.\ref{ribo}a) or sequestration of the ribosome binding site upon ligand binding (Fig.\ref{ribo}b). In order to assess the time scale in which such conformational change takes place and how it competes with ligand binding it is first important to quantitatively map the folding landscapes of riboswitches. From such landscapes  the  time scales for the conformational change in the switching region in the aptamer can be estimated \cite{Hyeon08PNAS,Lin08JACS}. 

In a pioneering experiment Block and coworkers used single molecule pulling experiments to map the folding landscape as a function of the extension of the RNA. Purine (guanine and adenine) riboswitches are remarkably selective in their affinity for ligands and carry out markedly different functions despite the structural similarity of their aptamers.  For the pbuE adenine (A) riboswitch, whose response to force was first probed in the LOT experiments, ligand binding activates the gene expression when an antiterminator is formed.  In the absence of adenine, part of the aptamer region is involved in the formation of a terminator stem with the expression platform resulting in transcription termination. The add A-riboswitch activates the gene expression by forming a translational activator upon ligand binding. In the absence of adenine, the riboswitch adopts the structure with a translational repressor stem in the downstream region.   At the heels of the first single molecule studies, we reported the entire folding landscape and calculated the time scale for switching of helix that engages in hairpin formation with the downstream sequence using the self-organized polymer (SOP) model \cite{Hyeon06Structure,HyeonBJ07,Lin08JACS}.  
As we show below comparison of the landscapes of these two riboswitches underscores the importance of the stability of the isolated helices in the assembly and rupture of the folded straucture.

\begin{figure}
\includegraphics[width=1.\columnwidth]{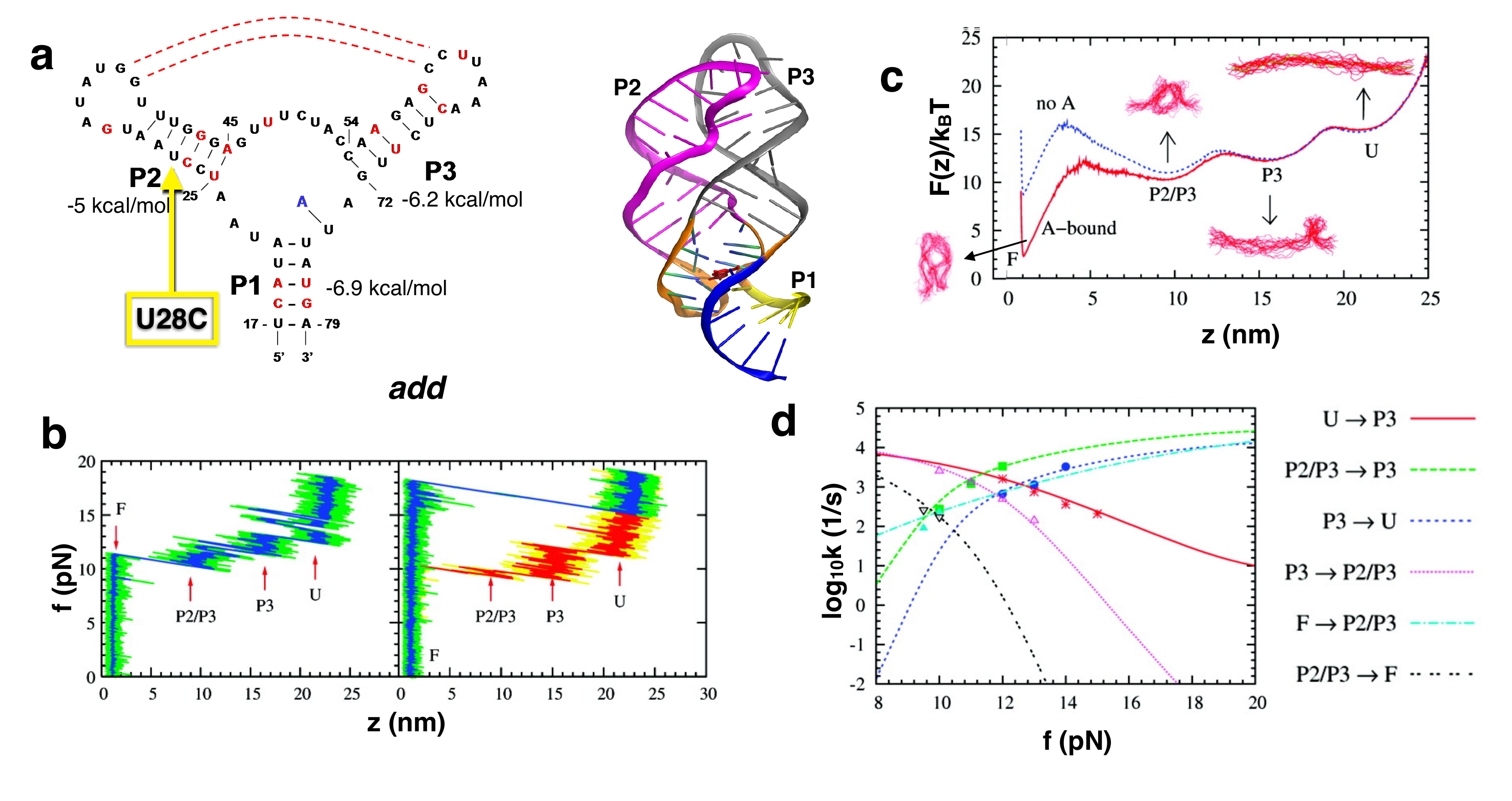}
\caption{Force-induced dynamics of add A-riboswitch (RS).
{\bf a.} Structure of the conserved domain of purine riboswitch containg a three way junction. On the left is the secondary structure map and on the right the three dimensional structure is shown. 
{\bf b.} Force-extension curves (FECs) obtained by pulling the RS at loading rate of 960 pN/s in the presence (left) and absence (right) of metabolite. 
The FEC in red on the right panel was obtained during the refolding of the RS while the exerted force is reduced. 
{\bf c.} Free energy profile $F(z)$ with (red) and without (blue) the metabolite. 
{\bf d.} Force-dependent transition rates. The data points are directly from simulation; The lines were obtained by calculating mean first passage time using $F(z)$ with a force-independent diffusion constant, which was calibrated by equating the theoretical and simulation rates.  
Figure adapted from Ref.\cite{Lin08JACS}
\label{add-RS}}
\end{figure}

Structures of purine riboswitch aptamers are characterized by a three-way junction consisting of P1, P2 and P3 helices, which are further stabilized by tertiary interactions in the folded state (Fig.\ref{add-RS}a). 
For pbuE A-riboswitch, binding of metabolite (adenine) activates the gene expression by enabling the riboswitch to form an antiterminator.
Without adenine, the molecule forms a terminator stem with the expression platform, resulting in transcription termination. 
On the other hand, the add A-riboswitch uses adenine to regulate the process of translation.
Recent single molecule experiments \cite{Greenleaf08Science,Neupane11NAR} and our simulation studies \cite{Lin08JACS,Lin14PCCP} have shown that, despite the marked structural similarity, these two aptamers have  different folding landscapes, thus providing a fingerprint of their function.

Single molecule optical tweezer experiments have been used to directly observe the hierarchical folding of both  pbuE A- and add A-riboswitch aptamers \cite{Greenleaf08Science,Neupane11NAR}. 
Here, we summarize force ($f$)-triggered unfolding and refolding of the A-riboswitch aptamer theoretically using Brownian dynamics simulations of the SOP model \cite{Hyeon06Structure,HyeonBJ07}. 
The crystal structure of add A-riboswitch (PDB id: 1Y26 (U17 to A79)) is available while that of pbuE A-riboswitch is not. 
However, since the sequence similarity between add-A and pbuE A-riboswitch is unusually high, we modeled the atomic structure of pbuE A-riboswitch by substituting the sequences of pbuE A-riboswitch into the crystal structure of add A-riboswitch and produced an ensemble of pbuE A-riboswitch structures via conformational sampling with molecular dynamics simulations \cite{Lin14PCCP}.

In the absence of adenine our simulation show that force-induced unfoldings of both pbuE-A and add A-riboswitches occur in three distinct steps.  
Force extension curves of riboswitch generated under constant loading condition ($r_f=960$ pN/s) 
reveal three distinct steps for both RS.  
Investigating the loss of secondary and tertiary contacts during the unfolding process, we found that the order of unfolding events differs qualitatively in add A-riboswitch and pbuE A-riboswitch. 
In add A-riboswitch, the unfolding occurred in the order of $\Delta$P1$\rightarrow$$\Delta$P2/P3$\rightarrow$$\Delta$P3$\rightarrow$U.
The order of forced unfolding of pbuE A-riboswitch is $\Delta$P1$\rightarrow$$\Delta$P2/P3$\rightarrow$$\Delta$P2$\rightarrow$U, where $\Delta$P2/P3 denotes the disruption of kissing loop interaction between P2 and P3 due to force.   
In the absence of adenine thermal fluctuations transiently disrupt this kissing-loop interaction, which is consistent with the observation that stable P2/P3 tertiary interactions require adenine.

The presence of adenine in the binding pocket in the triple-helix junction of add A-riboswitch changes the force-response of RS completely:
(i) The unfolding force increases from $\sim 10$ pN to $\sim 18$ pN, the value of which is comparable to the one found in experiments for the pbuE A-riboswitch aptamer \cite{Greenleaf08Science}; and (ii) the unfolding of RS occurs in all-or-none fashion without intermediate unfolding steps.  
After the complete unfolding, when refolding of the add A-riboswitch is initiated by reducing the force,  
we find that the refolding pathway follows the reverse order of unfolding pathway as U$\rightarrow$P3$\rightarrow$P2$\rightarrow$P2/P3$\rightarrow$P1.   
Refolding of P3 preceding that of P2 implies that P3 is more stable than P2, which is consistent with the implication from the stability of each helix ($\Delta G^{\text{add}}_{\text{P2}}=-5$ kcal/mol $>$ $\Delta G^{\text{add}}_{\text{P3}}=-6.2$ kcal/mol) calculated using the Vienna RNA package \cite{hofacker03NAR}.

Remarkably, despite the structural similarity between pbuE-A and add A-riboswitch aptamers, experiments show that P2 in pbuE unfolds at the last moment, which implies that P2 is the first structural element to refold upon force quench (or reduction). 
In agreement with the experiments, our results also imply that P2 ought to be more stable than P3 in the pbuE A-riboswitch aptamer, and Vienna RNA package indeed predicts that the stability of P2 is lower than that of P3 by 2 kcal/mol ($\Delta G^{\text{pbuE}}_{\text{P2}}=-7.3$ kcal/mol $<$ $\Delta G^{\text{pbuE}}_{\text{P3}}=-5.3$ kcal/mol) (Fig.3). 
The difference of the stability in the two RS aptamers explains the reversed order of the folding of P2 and P3 in the pbuE A-riboswitch aptamer. 
The order of unfolding of the helices, which is in accord with single molecule pulling experiments, is determined by the relative stabilities of the individual helices. Our results show that the stability of isolated helices determines the order of assembly and response to force in these non-coding regions. 
Thus, the folding landscape is determined by the local stability of the structural elements, a finding that also holds good in the thermal refolding of a number of RNA pseudoknots \cite{Cho09PNAS}. 

\begin{figure}
\includegraphics[width=1.\columnwidth]{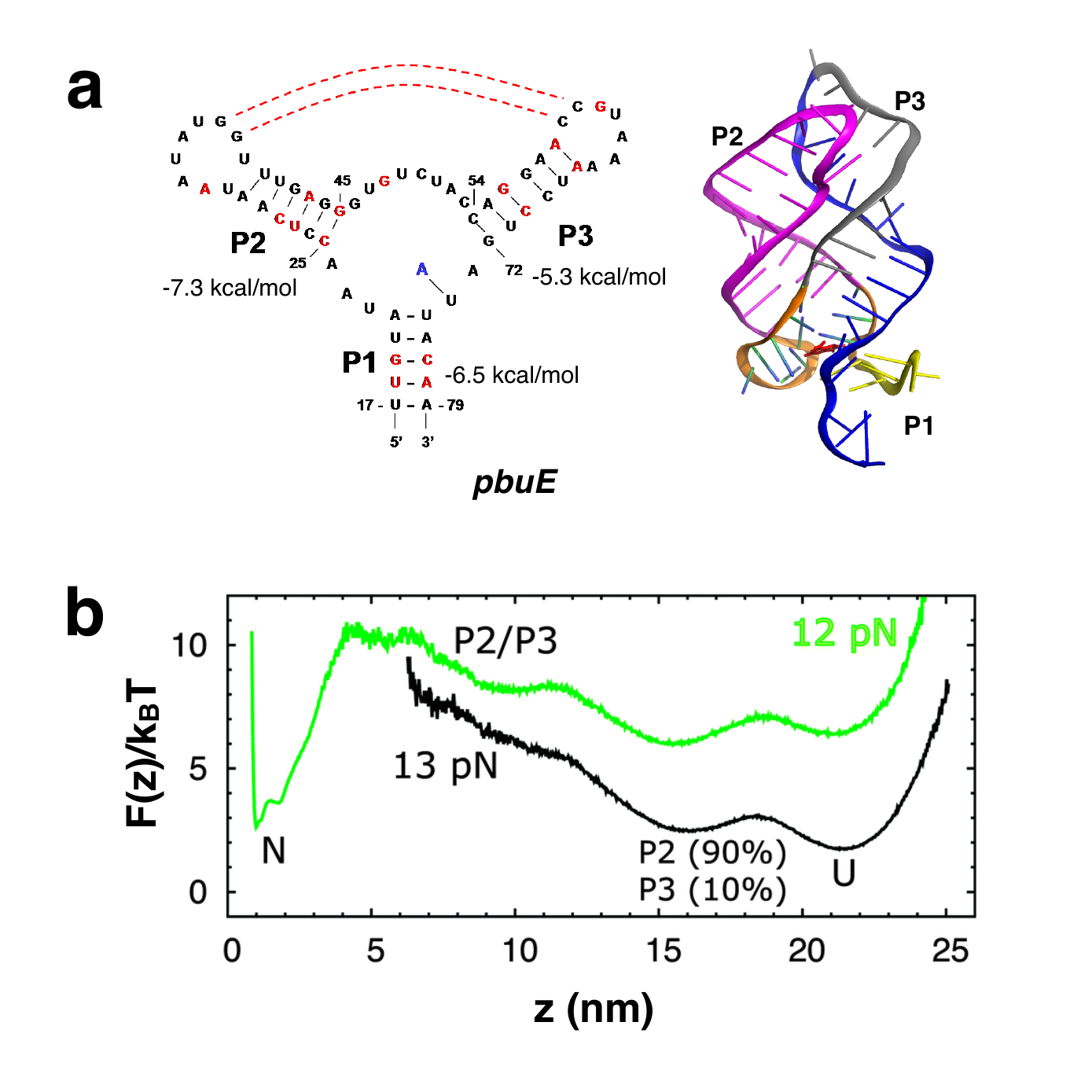}
\caption{pbuE adenine-riboswitch. 
{\bf a.} Secondary structure map (on the left) and the three dimensional structure on the right. 
{\bf b.} Free energy profile calculated from simulations at $f=12$, 13 pN. 
Figure adapted from Ref.\cite{Lin14PCCP}
\label{pbuE-RS}}
\end{figure}

Based on the stability hypothesis as the determining factor of the RNA folding landscape, we make an interesting prediction for pulling experiments in a mutant of the add A-riboswitch. 
One of the major differences that contributes the different stability of P2 helix in the two purine riboswitches is that the P2 of add A-riboswitch has one G-U and two G-C base pairs, whereas P2 in the pbuE A-riboswitch has three G-C base pairs (see Fig.\ref{add-RS}a and Fig.\ref{pbuE-RS}a). 
At the level of stability of secondary structure, a point mutation of U28C in the add A-riboswitch, which leads to three G-C base pairs in P2, would increase the secondary structure stability of P2 to $\sim$ 7.3 kcal/mol. 
Thus, the U28C mutation stabilizes the add A-riboswitch P2 by 1.1 kcal/mol lower than P3. 
The folding landscape of the U28C add A-riboswitch would be qualitatively similar to the WT pbuE A-riboswitch. 
As a consequence, we predict that U28C would reverse the order of unfolding of the add A-riboswitch. 

It is noteworthy that the number of contacts between P2 and P3 hairpin loops with and without adenine is almost identical, which suggests that adenine binding does not affect the interactions between P2 and P3 hairpin loops. 
Rather, the adenine binding stabilizes the triple-helix junction and makes unfolding of P1 more difficult. 
Hence, the rate limiting step in the fully folded aptamer is the formation of P1.

The free energy profiles ($G(z)$), obtained from the distribution of the molecular extension at force $f$ ($P(z;f)$) using $G(z;f) = -k_BT \log{P(z;f)}$, make explicit the hierarchical characteristics of RNA assembly and disassembly.  
In the absence of adenine, the position of first transition barrier from the folded state ($z = 1$ nm) is 
$\sim 2.5$ nm. 
Thus, the unfolding transition over the first barrier amounts to the unzipping of three base pairs in P1 helix. 
In the presence of adenine, the position of the first barrier shifts to $\sim4$ nm, 
implying that five base pairs of P1 in direct interaction with adenine should be disrupted at the first TS. 
Thus, adenine binding makes the first unfolding barrier the rate limiting step. 
We also find that, at $f = 10$ pN, binding of adenine stabilizes the folded state by $\sim$ 6 $k_BT$ and increases the energy barrier for leaving the folded state by 2 $k_BT$. 
These results support the hypothesis that the ligand binding stabilizes P1, a feature that is common to most riboswitches. 

The binding of adenine stabilizes the folded basin of attraction, which slows down the unfolding transition by two orders of magnitude. 
The slow unfolding rate, which make the conformational sampling difficult, is in agreement with that observed for FMN riboswitches \cite{Wickiser2005MolCell}, suggesting that functions of riboswitches are kinetically, rather than thermodynamically, controlled (see below for further discussion). 
The result is crucial for transcription of the complete riboswitch.

\section{Folding landscapes of SAM riboswitch. Is SAM riboswitch under thermodynamic control?}

\begin{figure}
\includegraphics[width=1.\columnwidth]{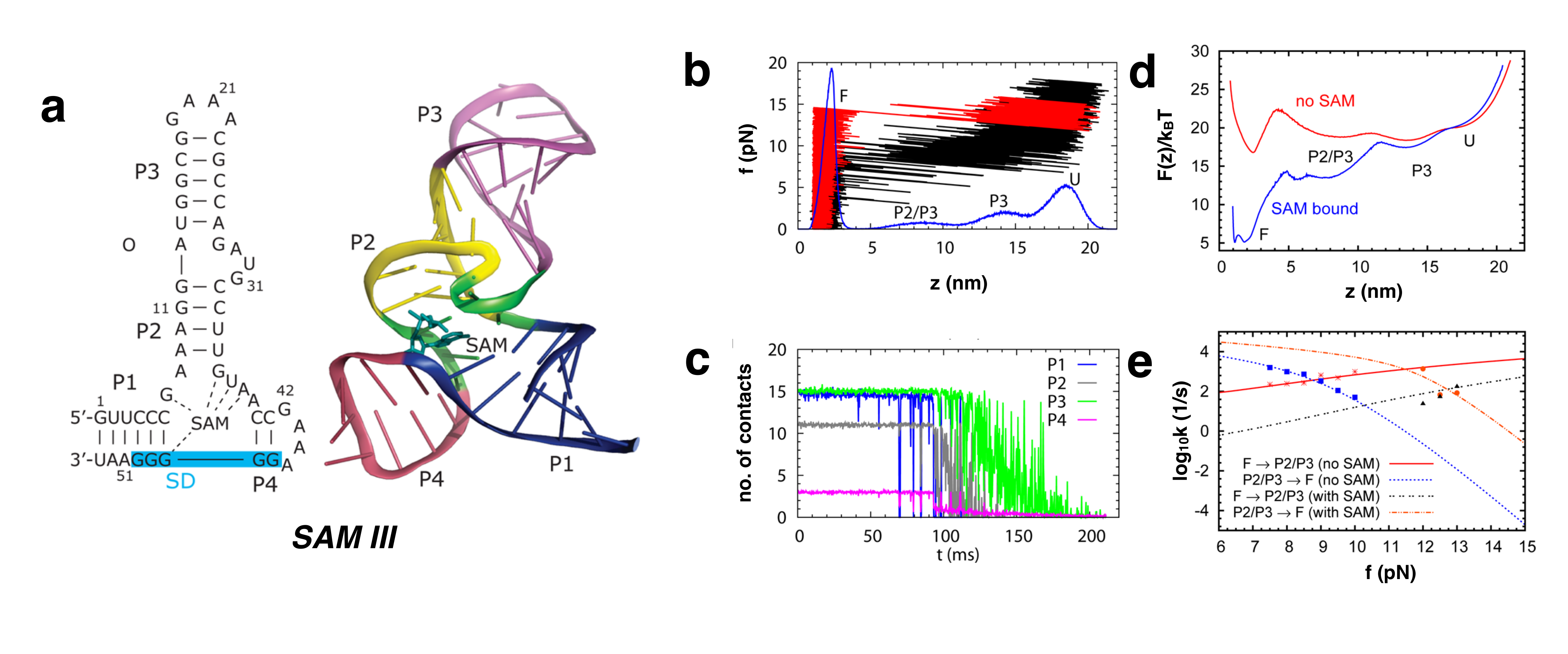}
\caption{Dynamics of SAM-III riboswitch under force. 
{\bf a.} Structure of SAM-III RS.  The blue shaded area on the left indicates the Shine-Dalgarno sequence recognized by the ribosome.
{\bf b.} Simulated force-extension curve of SAM-III riboswitch in the absence of metabolite (black) produced at $r_f=96$ pN/s. 
The distribution of molecular extension ($z$) during the pulling simulation is shown in blue at the bottom. 
FEC in red was produced in the presence of metabolite at the binding pocket. 
{\bf c.} Average number of contacts in each helix from P1 to P4. 
{\bf d.} Free energy profile at zero force calculated from streching simulation with and without metabolite (SAM) in the binding pocket. 
{\bf e.} Transition rates between F and P2/P3 states at varying forces. The data points are from explicit simulations. The lines were obtained from mean first passage time calculation on $F(z)$. 
Figure adapted from Ref.\cite{lin2013JACS}
\label{SAM-RS}}
\end{figure}

Upon binding S-adenosylmethionine (SAM) the riboswitch undergoes 
undergoes an allosteric transition to control translation. There are 
at least five distinct classes of riboswitches that bind  
SAM or its derivative S-adenosylhomocysteine (SAH). One of them, the SAM-III riboswitch \cite{Fuchs2006NSMB} in the metK gene 
(encodes SAM synthetase) from {\it Lactobacillales} species, inhibits 
translation by sequestering the Shine-Dalgarno (SD) sequence (Fig.\ref{SAM-RS}a)  \cite{Fuchs2006NSMB}. When SAM is bound, the somewhat atypical 
SD sequence (GGGGG shown in the blue shaded area in Fig.\ref{SAM-RS}a) is sequestered by 
base pairing with the anti-Shine-Dalgarno (ASD) sequence, thus hindering the 
binding of the 30S ribosomal subunits to mRNA. 
In the absence of SAM, the sequence outside the
binding domain  enables the riboswitch
 to adopt an alternative folding pattern, in which the SD
sequence is exposed and free to engage the ribosomal subunit \cite{Lu2011JMB}. 
Thus, SAM-III is an OFF switch for translation, in contrast to 
{\it add} A-riboswitch, which is an ON switch. 
Translation control is determined by competition between the SD-ASD 
pairing and loading of ribosomal subunit onto the SD sequence.
In order to function as a switch, the SD sequence has to be exposed for ribosome recognition, which implies
that at least part of the riboswitch structure accommodating SAM III has to unfold (Fig.\ref{SAM-RS}). These considerations prompted us to 
 quantitatively determine the 
folding landscape and the rates of conformational transitions between the ON and OFF states of the SAM-III riboswitch \cite{Anthony2012PNAS}.

In order to understand if SAM-III functions under thermodynamic control we calculated the force-extension ($f,z$) or FEC as well as constant $f$-dependent free energy profiles as a function of $z$ for SAM-III.  The FEC (black curve in Fig.\ref{SAM-RS}b) 
shows that there are two intermediate states, 
with extension $z \sim 14$ nm and $z \sim 9$ nm in the absence of SAM.  
Helices P1 and P4 rupture in a single step at $\sim$ 9 pN and P2 unravels at $f\approx 10-11$ pN. Helix P3 unfolds fully only
when $f >$ 12pN. Interestingly, when SAM is bound, the riboswitch
unfolds in an apparent all-or-none manner at $f \sim 15$ pN (Fig.\ref{SAM-RS}b).
The distribution $z$ (blue curve in Fig. 5A), shows
the presence of two intermediate states ahead of global unfolding. The 
$z \sim 9$ nm peak corresponds to rupture of P1 and P4 helices.
 P3 unfolds in the later stages creating a peak at 
$z \sim 14$ nm. The
hierarchical unfolding pathway of SAM-III riboswitch
is $F \rightarrow \Delta \text{P1} \Delta \text{P4} \rightarrow \Delta \text{P3} \rightarrow \text{U}$,
where $\Delta \text{P1} \Delta \text{P4}$ means helices P1 and P4 are ruptured, and 
$\Delta$P2 represents additional unfolding of P2.  The observed order of unfolding is also reflected in the rupture
of contacts, a more microscopic representation of unfolding dynamics (Fig. \ref{SAM-RS}c). The intermediate states in Fig.\ref{SAM-RS}b can be
traced to the breaking of contacts within the helices. Just as in Fig. 5b the order of contact rupture corresponds to the order in which the helices unfold in the FEC in Fig.\ref{SAM-RS}b. 

Using simulations at constant force we also calculated the free energy profiles  using
$F(z,f) = -k_{B}T\log{P(z)}$ where $P(z)$ is the probability distribution of $z$. At $f < 9$ pN and in the absence of SAM the riboswitch is in the folded basin of attraction (Fig. 5d).  
The free energy profile in Fig. 5d also shows that binding of SAM consolidates
the formation of helix P1 and P4, further stabilizing the folded state. 
At $f = 9$ pN, SAM binding stabilizes the folded 
state by $\sim 12$ $k_{B}T$, and increases the energy 
barrier for leaving the folded state by $\sim 3$ $k_{B}T$.
The distance from the folded 
state to the first barrier in the absence of SAM is $\sim 2$ nm, 
which indicates unzipping of 2.5 base pairs, assuming a contour length increase of
0.4 nm/nt.  In the presence of SAM, the position of the first barrier 
shifts to $\sim 5$ nm, implying that 4 base pairs of P1 next to the 
nucleotide G48 (Fig.\ref{SAM-RS}a) that has direct contacts with SAM are
ruptured at the transition state. Thus, disruption of contacts with SAM becomes
the key barrier in the first unfolding step, and must be an important step  in translational regulation.

\section{Is SAM riboswitch under thermodynamic control?}

As shown in Fig.\ref{RSkineticnetwork}a, the kinetic processes in riboswitches that control transcription are determined by a number of time scales.  
In the transcription process, the ability to function as an efficient switch depends on an interplay of the time scales:  
(i) metabolite binding rate ($k_b$), (ii) the folding times of the aptamer ($k_f$), (iii) the time scales to switch and adopt alternate conformations with the downstream expression platform ($k_t$), and (iv) the rate of transcription.  
In ``OFF" riboswitches that shut down gene expression upon metabolite binding, a decision to terminate transcription has to be made before the terminator is synthesized, which puts bounds on the metabolite concentration, and the aptamer folding rate ($k_f$). 
For simplicity, $\gamma=k_t/k_f$ can serve as a simple criterion to determine whether the co-transcriptional folding of riboswitches is under thermodynamic or kinetic control. 
In the limit $\gamma\gg1$ transcript synthesis is faster than the equilibration time of the riboswitch conformation. 
For typical values of these parameters in both FMN and pbuE A-riboswitch, efficient function mandates that the riboswitches be under kinetic control, which implies that the ``OFF" and ``ON" states of riboswitch are not in equilibrium. 

In contrast, the function of SAM-III, which controls translation, is different. 
The major time scales that control the function of SAM-III RS, and those that regulate translation in general, are 
(i) bimolecular binding rate of SAM to RS ($k_b$), 
(ii) dissociation rate of SAM from the riboswitch complex ($k_{-b}$),   
(iii) the rate of mRNA degradation ($k_{\text{mRNA}}$).  
Thus, the only clear physical bound on the function of SAM-III is that binding of metabolite should occur multiple times before the mRNA degradation, which leads to $k_b[M]\gg k_{\text{mRNA}}$, where $[M]$ is the concentration of SAM. 
Typical values of $k_b\sim 0.11$ $\mu  M^{-1}s^{-1}$, $k_{mRNA}\approx 3$ $min^{-1}$ and $k_{dis}\approx 0.089$ $s^{-1}$ requires that $[M]\gtrsim 50$ nM. 
It is worth pointing out that our estimates of folding and unfolding times based on simulations at low forces, and other time scales are all much less than $k_{\text{mRNA}}^{-1}\approx 20$ sec, which sets the longest time for translational control. 
Hence, the multiple transitions between the OFF and ON states can occur before mRNA is degraded, which gives additional credence to the argument that the function of SAM-III is under thermodynamic control. There is a caveat to this conclusion. It is known that in bacteria transcription and translation are coupled, which is likely to complicate our arguments. In order to provide a complicate description we require a network model that includes transcription-translation coupling. Because $k_{\text{mRNA}}$ is small it is still possible that the SAM riboswitch could be under thermodynamic control.

\section{Kinetic Network Model of gene regulation and the role of negative feedback in control of transcription}
\begin{figure}
\includegraphics[width=0.65\columnwidth]{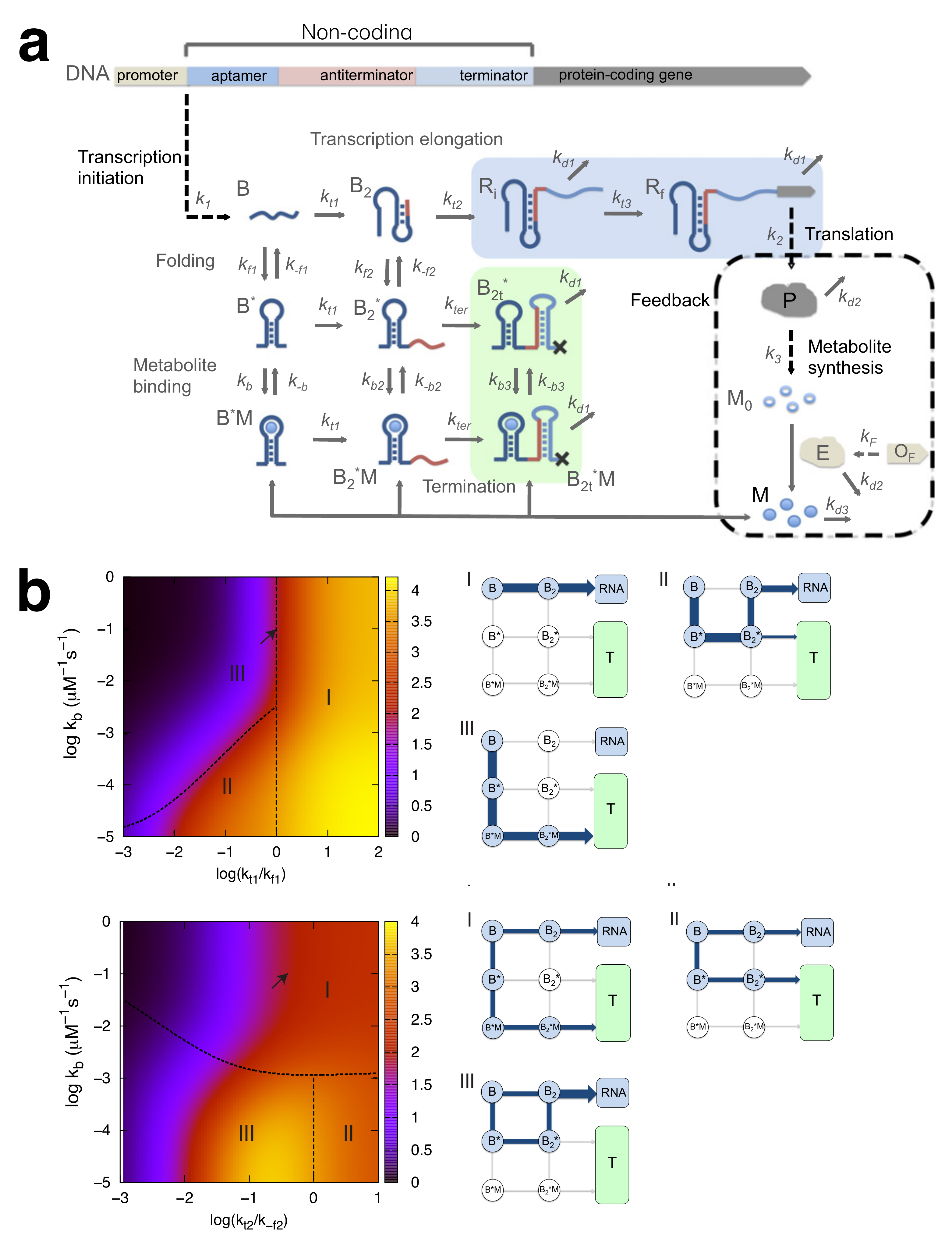}
\caption{{\bf a.} Kinetic network model for transcription regulated by ``OFF"-riboswitches. 
{\bf b.} Dependence of protein production on the network parameters with ``negative feedback" ($k_{f1}=0.1$ $s^{-1}$, $k_{-f1}=0.04$ $s^{-1}$, $k_{f2}=2.5\times 10^{-3}$ $s^{-1}$, $k_{-f2}=0.04$ $s^{-1}$, $k_{t1}=0.1$ $s^{-1}$, $k_{t2}=0.016$ $s^{-1}$, $k_b=0.1$ $\mu M^{-1}s^{-1}$, $k_{-b}=10^{-3}$ $s^{-1}$, $k_{t3}=0.01$ $s^{-1}$, $K_1=0.016$, $k_2=0.3$ $s^{-1}$, $k_3=0.064$ $s^{-1}$, $k_{d1}=2.3\times10^{-3}$ $s^{-1}$, $k_{d2}=2.7\times 10^{-4}$ $s^{-1}$, $k_{d3}=4.5\times 10^{-3}$ $s^{-1}$, and $\mu=5\times 10^{-4}$ $s^{-1}$.   
(Top) Protein levels $[P]$ (color coded) as functions of $k_{t1}/k_{f1}$ and $k_b$). 
The dependence of $[P]$ on $k_{t1}$ and $k_b$ is categorized into three regimes. 
Points on the dashed line separating regime II and regime III satisfy $k_b[M]=k_{t1}$. 
The major pathway in the transcription process in each regime is shown on the right. 
The arrow indicates the data point from the value of $k_{t1}=0.1$ $s^{-1}$ and $k_b=0.1$ $s^{-1}$. 
(Bottom) $[P]$ as functions of $k_{t2}/k_{f2}$ and $k_b$. 
Points on the dashed line separating regime I and II/III satisfy $k_{b1}[M]=k_{-f2}$. 
The data point corresponding to the arrow results from using the value of $k_{t2}=0.016$ $s^{-1}$ and $k_b=0.1$ $s^{-1}$.
Figure adapted from Ref.\cite{lin2012BJ}
\label{RSkineticnetwork}}
\end{figure}

As stated earlier, gene expression is mediated by binding of metabolites to the conserved aptamer domain, which triggers an allosteric reaction in the downstream expression platform. However, the target metabolites are usually the products or their derivatives of the
downstream gene that the riboswitches control. Hence, metabolite binding to
riboswitches serves as a feedback signal to control RNA transcription or
translation initiation.
The feedback through metabolite binding is naturally designed to be 
a fundamental network motif for riboswitches. 
In ON-riboswitches, metabolite binding thus stabilizes the aptamer structure during transcription and prevents the formation of the terminator stem before transcription is completed (pbuE A-riboswitch) or the formation of translation repressor stem before translation is initiated (add A-riboswitch).
Whereas in OFF-riboswitch, metabolite binding shuts down the gene expression by promoting the formation of terminator stem (see Fig.\ref{RSkineticnetwork}).  In order to understand the {\it in vivo} riboswitch we developed a kinetic network model
taking into account the interplay between the
speed of RNA transcription, folding kinetics of the nascent RNA 
transcript, and the kinetics of metabolite binding to the nascent RNA 
transcript, and the role of feedback arising from interactions 
between synthesized metabolities and the transcript.
The effects of speed of RNA transcription and metabolite binding kinetics have also been 
investigated experimentally {\it in vitro} in an insightful study involving the flavin 
mononucleotide (FMN)
riboswitches \citep{Wickiser2005MolCell}. They argued that FMN riboswitch is kinetically 
driven implying that the riboswitch does not reach thermodynamic equilibrium 
with FMN before a decision between continued transcription and 
transcription termination needs to be made. The mathematical solution of the kinetic network model, which uses as partial input the rates of switching obtained from the folding landscapes, show that in general riboswitches that control transcription are under kinetic control \cite{lin2012BJ}. A brief summary is presented here.

Efficient function of RS, implying a large dynamic range (quantified by response of the RS to varying metabolite concentration) without compromising the requirement to suppress transcription or translation, is determined by a balance between the transcription speed ($k_{\text{trxn}}$), the folding and unfolding rates of the apatmer ($k_f$ and $k_{-f}$), and the binding and unbinding rates of the metabolite ($k_b[M]$ and $k_{-b}$, where $[M]$ is the metabolite concentration).  In order to capture the physics behind the dynamics it is necessary to consider kinetic network model describing the coupling between aptamer dynamics and transcription.
In Fig.\ref{RSkineticnetwork} demonstrating the kinetic network model for the transcription regulation by OFF-riboswitches, the upstream of the protein-coding gene consists of sequences involving the transcriptions of aptamer (B), antiterminator (B$_2$) and terminator (B$_2^*$) of the riboswitch. 
The transcription initiation is followed by elongation, folding of the RNA transcript, and metabolite binding. 
RNA polymerase first transcribes the aptamer (B), and moves on to the synthesis of the RNA transcript for anti terminator B$_2$ at a rate of $k_{t1}$, and terminator sequence at $k_{t2}$, resulting in the production of the regulatory region of RNA.  
$R_i$ is the transcript with the sequence of the protein-coding region starting to be transcribed, and eventually grows to $R_f$, the full protein-coding region transcribed, with a rate of $k_{t3}$. 
During the process of transcription elongation, each of the transcript states, B and B$_2$, can form states with the aptamer domain folded (B$^*$ and B$^*_2$) with a folding rate of $k_{f1}$ and $k_{f2}$, respectively. 
The folded aptamers bound with metabolite (M) are B$^*$M and B$^*_2$M with binding rate constant $k_b$ and $k_{b2}$, respectively. 
The transcripts in state B$^*_2$ and B$^*_2$M can further elongate until the terminator sequence is transcribed with their expression platform forming a transcription terminator stem and dissociate from the DNA template with a rate of $k_{ter}$, forming B$^*_{2t}$ and B$^*_{2t}$M. 
The fraction of transcription termination, $f_{ter}$, is determined from the amount of the terminated transcripts (in green block) relative to nonterminated transcripts (in blue block). 
The activated metabolte (M), produced from protein $P$ and activated by the enzyme ($E$) encoded by the gene OF, can bind to the folded aptamer and can abort transcription, which imposes a negative feedback on the transcription process.  

For a riboswitch to function with a large dynamic range, transcription levels should change significantly as the [M] increases from a low to high value. 
(i) In the high [M], 
RNA transcript in the aptamer folded state binds a metabolite with $k_b[M]$. 
In FNM-riboswitch, small $k_{-b}$ value results in the formation of a terminator stem, which subsequently terminates transcription. 
(ii) In the low [M] limit, the aptamer folded state is mostly unbound and can remain folded until transcription termination or can fold to the antiterminator state, enabling the synthesis of full RNA transcript. 
The levels of transcription termination are thus controlled by the transition rates between the aptamer folded and unfolded states ($k_{f1}$, $k_{-f1}$, $k_{f2}$, $k_{-f2}$).
Equilibrium between B$_2$ and B$^*_2$ can be reached only if the rate of transcription is much slower than the rates of folding/unfolding and metabolite binding. 
By varying the rate of transcription, which can be experimentally realized by adding transcription factors such as NusA \cite{Zhou2011MolecularCell}, it may be possible to drive the cotranscriptional folding of riboswitch from kinetic to thermodynamic control. However, for realistic values of the various rates in Fig.\ref{RSkineticnetwork} we predict that transcription {\it in vivo} is under kinetic control.

In the presence of a negative feedback loop the concentration of target metabolites is also regulated by gene expression. Under nominal operating conditions 
($\gamma_2=k_{t2}/k_{-f2}\sim 0.01-0.1$) binding of target metabolites, products of the downstream gene that riboswitches regulate, significantly suppresses the expression of proteins. Negative feedback suppresses the protein level by about half relative to the case without feedback. In vivo, the presence of RNA binding proteins, such as NusA \cite{Zhou2011MolecularCell}, may increase the pausing times, thus effectively reducing the transcription rates. 
Thus, the repression of the protein level by the riboswitch through metabolite binding may be up to 10-fold. Faster RNA folding and unfolding rates than those we obtained may also increase the suppression by negative feedback and broaden the range of transcription rates over which maximal suppression occurs. These predictions are amenable to experimental test.

In response to changes in the active operon level, the negative feedback speeds up the response time of expression and modestly reduces the percentage change in the protein level relative to change in the operon level. The steady-state level of expression for autoregulation varies as a square root of the DNA concentration. Adaptive biological systems may minimize the variation in gene expression to keep the systems functioning normally even when the environments change drastically. One may need to consider more complex networks than the single autoregulation in the transcription network to find near perfect adaptation to the environmental change \cite{Ma2009Cell}.

The effect of negative feedback accounting for the binding of metabolites, which themselves are the product of genes that are being regulated. Our previous work showed that because of the interplay of a number of time scales determining the riboswitch function at the system there are many scenarios that can emerge, which can be encapsulated in terms of a dynamic phase diagram. 
An example dynamic phase diagram (for a full discussion see \cite{lin2012BJ}) in terms of  the transcription rates $k_{\text{trxn}}$, $k_f$, $k_{-f}$, $k_b[M]$ illustrates the complexity of the transcription process. The interplay between
folding of RNA transcripts, transcription, and metabolite binding regulate
the expression of $P$, which can be quantified using the production of the protein,
$[P]$,  on the transcription rates and the effective binding rate $k_{b}[M]$.
The dynamic phase diagram in Fig. 4B, calculated by varying both $k_{t1}$ 
($k_{t2}$) and $k_{b}$ with the equilibrium binding constant of the metabolite to the aptamer fixed to $K_{D}=10$ nM, a value that is appropriate for FMN \cite{Wickiser2005MolCell}. 
We expect that after the aptamer
sequence is transcribed, the formation of the aptamer structure is the key 
step in regulating transcription termination. Thus,  regulation of $[P]$ should be controlled by the folding rate $k_{f1}$
,the effective metabolite binding rate, and  $k_{t1}$ for regulation of $[P]$. Figure 4B shows three regimes for the
dependence of $[P]$ on $k_{t1}$ and $k_{b}[M]$. In regime I, $k_{t1} > k_{f1}$,
the folding rate is slow relative to transcription to the next stage (Fig. 1), which implies that 
the aptamer structure does not form on the time scale set by transcription. The dominant 
flux  is from
$B$ to $B_{2}$, which leads to high probability of fully transcribed RNA
downstream because of the low transition rate from $B_{2}$ to $B_{2}^{*}$.
The metabolite binding has little effect on protein expression in this regime, 
particularly for large $k_{t1}/k_{f1}$, and hence the protein is highly 
expressed. In regime II, $k_{b}[M] < k_{t1} < k_{f1}$, the aptamer has enough 
time to fold but metabolite binding is slow. The dominant flux is 
$B\rightarrow B^{*}\rightarrow B_{2}^{*}$, leading to formation of 
antiterminator stem ($B_{2}^{*}\rightarrow B_{2}$) or transcription 
termination ($B_{2}^{*}\rightarrow B_{2t}^{*}$). The expression level of 
protein is thus mainly determined by $k_{-f2}$ and $k_{t2}$, and the protein
production is partially suppressed in this regime. 
In regime III, $k_{t1} < k_{f1}$ and $k_{t1} < k_{b}[M]$, the aptamer has 
sufficient time to both fold and bind metabolite, the dominant pathway is 
$B\rightarrow B^{*} \rightarrow B^{*}M \rightarrow B_{2}^{*}M$, leading to
transcription termination. The protein production is highly suppressed in this
regime. The arrow shows that for parameters that are appropriate for FMN riboswitch (see Tables 1 and Table 2 in \cite{lin2012BJ}($k_{t1}$ fall 
on the interface of regime I and regime III. The metabolite binding fails to reach thermodynamic equilibrium due to low
dissociation constant. However, the effective binding rate is high because the 
steady state concentration of metabolites ($\sim$ 25 $\mu$M) is in large 
excess over RNA transcripts. Thus, the riboswitch is kinetically driven under 
this condition even when feedback is included.

\section{Concluding Remarks}
Based on our previous works, we have provided broad overview, from atomic scale to systems level, of  how the complex dynamics of riboswtiches emerge depending many inter-related rates. 
At the atomic scale, dynamics of the surface water around riboswitches plays critical role in inducing the local fluctuation in the riboswtich.  
At the level of single riboswitch, we have shown that explicit simulations of riboswitches, in conjunction with single molecule experiment, is a powerful tool to understand the conformational dynamics of riboswtich both with and without metabolites. 
At the systems level, in which the minimal model of cellular environment is considered, 
the dynamics of riboswitches in isolation are modulated by a number of factors, which is made explicit by the kinetic network model of FMN riboswitch.     Our collective works show that combination of theory, experiments, and simulations are needed to understand the function of riboswitches under cellular conditions.

Riboswitches also provide novel ways to engineer biological circuits to control gene expression by binding small molecules. 
As found in tandem riboswitches \cite{Breaker08Science,Sudarsan2006Science}, multiple riboswitches can be engineered to control a single gene with greater regulatory complexity or increase in the dynamic range of gene control. Synthetic riboswitches have been successfully used to control the chemotaxis of bacteria \cite{Topp2007JACS}. Our study provides a physical basis for not only analyzing future experiments but also in anticipating their outcomes.

{\bf Acknowledgements:} This work was supported by grants from the National Institutes of Health (GM 089685) and the National Science Foundation (CHE 13-61946).

\bibliographystyle{apsrev}
\bibliography{mybib1,ref}
\clearpage 

\end{document}